\documentclass[twocolumn]{aa} 

\usepackage{graphicx}
\usepackage{txfonts}

\begin{document} 

   \title{Magnetohydrostatic modeling of AR11768 based on a \textsc{Sunrise}/IMaX vector magnetogram}

   \author{X. Zhu\inst{1}, T. Wiegelmann\inst{1}, and S. K. Solanki\inst{1,2}}

  \institute{Max-Planck-Institut f\"{u}r Sonnensystemforschung, Justus-von-Liebig-Weg 3, 37077 G\"{o}ttingen, Germany\\
            \email{zhu@mps.mpg.de}
            \and
            School of Space Research, Kyung Hee University, Yongin, Gyeonggi 446-701, Republic of Korea}

   \date{Received ; accepted }

  \abstract
   {High resolution magnetic field measurements are routinely done only in the solar photosphere. Higher layers like the chromosphere and corona can be modeled by extrapolating these photospheric magnetic field vectors upward. In the solar corona, plasma forces can be neglected and the Lorentz force vanishes. This is not the case in the upper photosphere and chromosphere where magnetic and non-magnetic forces are equally important. One way to deal with this problem is to compute the plasma and magnetic field self-consistently, in lowest order with a magnetohydrostatic (MHS) model. The non-force-free layer is rather thin and MHS models require high resolution photospheric magnetic field measurements as the lower boundary condition.}
   {We aim to derive the magnetic field, plasma pressure and density of AR11768 by applying the newly developed extrapolation technique to the \textsc{Sunrise}/IMaX data embedded in SDO/HMI magnetogram.}
   {An optimization method is used for the MHS modeling. The initial conditions consist of a nonlinear force-free field (NLFFF) and a gravity-stratified atmosphere. During the optimization procedure, the magnetic field, plasma pressure and density are computed self-consistently.}
   {In the non-force-free layer, which is spatially resolved by the new code, Lorentz forces are effectively balanced by the gas pressure gradient force and the gravity force. The pressure and density are depleted in strong field regions, which is consistent with observations. Denser plasma, however, is also observed at some parts of the active region edges. In the chromosphere, the fibril-like plasma structures trace the magnetic field nicely. Bright points in \textsc{Sunrise}/SuFI 3000 $\AA$ images are often accompanied by the plasma pressure and electric current concentrations. In addition, the average of angle between MHS field lines and the selected chromospheric fibrils is $11.8^\circ$, which is smaller than those computed from the NLFFF model ($15.7^\circ$) and linear MHS model ($20.9^\circ$). This indicates that the MHS solution provides a better representation of the magnetic field in the chromosphere.}
   {}


\titlerunning{MHS extrapolation application to IMaX data}
   \maketitle

\section{Introduction} \label{sec:intro}

Due to the high plasma $\beta$ in the photosphere and chromosphere, non-magnetic forces should be taken into account when we study these layers. If we neglect dynamics and plasma flow, then the resulting static state of the system can be described by  the so-called magnetohydrostatic (MHS) assumption, which is determined by the balance of Lorentz force, pressure gradient and gravity force, together with the solenoidal condition of the magnetic field.

The 3D MHS equations can be linearized by some assumptions. The analytical solutions for the linear MHS (LMHS) model have been developed in many papers \citep[e.g.,][]{l85,l91,o85,n97,pn00,nw19}. These solutions have been used for a number of specific applications to the Sun \citep[e.g.,][]{bg91,zh93,ads98,adm99,p00,rwi08,gfm13,wnn15,wnn17} and stars \citep{la08,mgn16}.

A more challenging problem is to solve the MHS equations in the generic case, which can be done only numerically. A few methods have been developed. The magnetohydrodynamic (MHD) relaxation method was introduced to derive the MHS solution by using an ``evolution technique'' \citep{mm94,jmm97,zwd13,zwd16,mki19}. The Grad-Rubin iteration, which is well known in the calculation of the nonlinear force-free field (NLFFF), was extended by \cite{gw13} and \cite{gbb16} to solve the MHS equations. \cite{wi03} and \cite{wn06,wnr07} develop the optimization method to treat the MHS equations. We recently extended this method by introducing the gravity force \citep{zw18}. In addition, our new algorithm ensures that the resulting plasma pressure and density are positive definite. More recently, we further test our code with the radiative MHD simulation of a solar flare \citep{zw19}. This challenging test (solar flares are intrinsically non-static and even non-stationary) provides a solid foundation for the application of the code to real observations.

It is worth noting that, besides the traditional NLFFF models, a new type of NLFFF model \citep{msd12,a13,a16} has been introduced recently. This alternative approach uses the line-of-sight magnetogram to constrain the potential field. With the help of a forward-fitting method the angle between the magnetic field and coronal or chromospheric loops is minimzed. In this model, the assumption of a force-free photosphere is not used either. This works well in the corona, but may not be the ideal way to describe the non-force-free chromospheric field.

In this paper, we apply our code, for the first time, to a vector magnetogram obtained by the IMaX instrument on the \textsc{Sunrise} balloon-borne solar observatory. Since the field-of-view (FOV) of IMaX is limited to part of the active region, an SDO/HMI vector magnetogram is used to cover the unobserved parts. The organization of the paper is as follows. In Section \ref{sec:method}, we describe the dataset used in this paper. In Section \ref{sec:results}, we analyze the results and compare them with other models. Conclusions and perspectives are presented in Section \ref{sec:conclusion}.

\section{Magnetohydrostatic Extrapolation and its application to AR11768} \label{sec:method}

The MHS extrapolation computes the magnetic field, plasma pressure and density consistently with the help of an optimization principle. The algorithm was described and tested in detail in \cite{zw18,zw19}.

\begin{figure}
\includegraphics[width=\hsize]{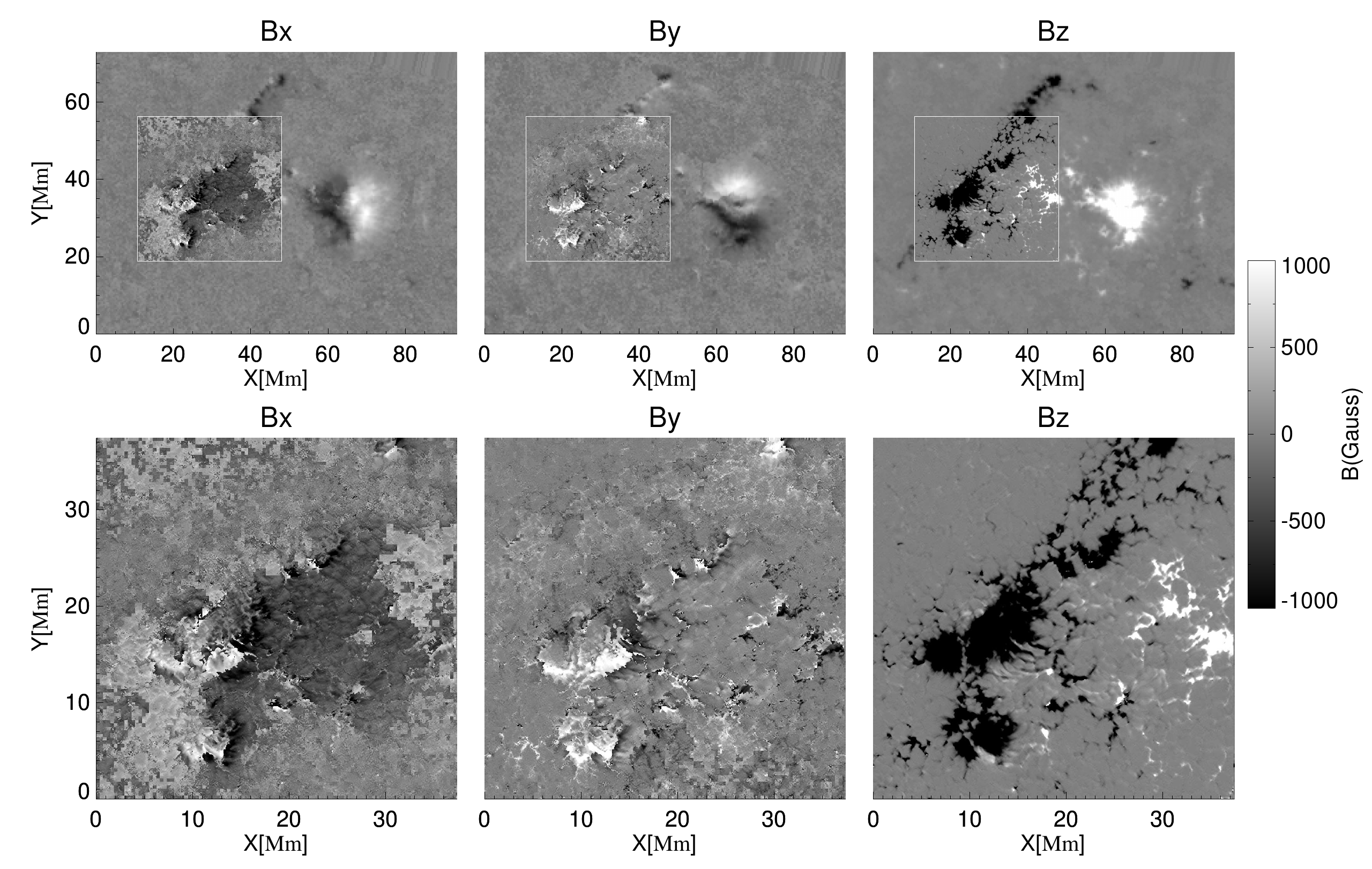}
\caption{Top: IMaX measurements at 23:48 UT embedded in the vector magnetogram of HMI (observed at 23:48 UT). The outlined region with a clearly visible higher resolution is the IMaX FOV. Bottom: vector magnetogram of IMaX.}
\label{fig:magnetogram}
\end{figure}

The primary dataset used in this work was recorded by the vector magnetograph IMaX \citep{mda11} onboard the \textsc{Sunrise} balloon-borne solar observatory \citep{bgs11,bss11,ggb11} during its second flight, refer to as \textsc{Sunrise} II \citep{srb17}. The IMaX data have a pixel size of 40 km a FOV that contains $936\times936$ pixels ($50''\times 50''$). This FOV is limited to part of AR11768. The data were inverted by \cite{krs17} using the SPINOR inversion code \citep{fsf00} that builds on STOPRO routines \citep{s87}. A one-component atmospheric model with three optical depth nodes (at log$\tau$=-2.5, -0.9, 0) for the temperature and height-independent magnetic field is applied. The affect of the inversion (not the same inversion used here) on the extrapolation result was studied by \cite{wys10} and found to be minor. The 180$^\circ$ ambiguity is removed with an acute angle method \citep{mlb06} which minimizes the angle with, in this case, the corresponding HMI vector magnetogram. We note that there is a square pattern in the transverse field of IMaX (Fig.\ref{fig:magnetogram}). It originates from disambiguating IMaX with HMI data in which a square patten also exists (seen in the transverse field). Essentially, the square pattern is the IMaX noise that is modulated to match the HMI spatial resolution by disambiguating. Although this noise appears in big mosaics (big compared to the IMaX resolution), the size of every single one is still very small compared to the active region. Therefore the effect they have on the extrapolation is expected to be similar to the effect of the normal noise which has been studied by \cite{zw18}. In that paper we found, based on the quantitative assessment for the extrapolation, the influence of the random noise (20\% level) on the magnetic field to be less than 4\%. As pointed out by \cite{dsb09}, it is necessary to have flux-balance within the FOV and to catch the magnetic connectivity in order to find unique solutions. So we have embedded the IMaX data into the SDO/HMI vector magnetogram \citep{ptc12,ssb12} taken closest in time to the analyzed IMaX magnetogram. Fig.~\ref{fig:magnetogram} shows both IMaX and combined vector magnetogram.

\begin{figure}
\includegraphics[width=\hsize]{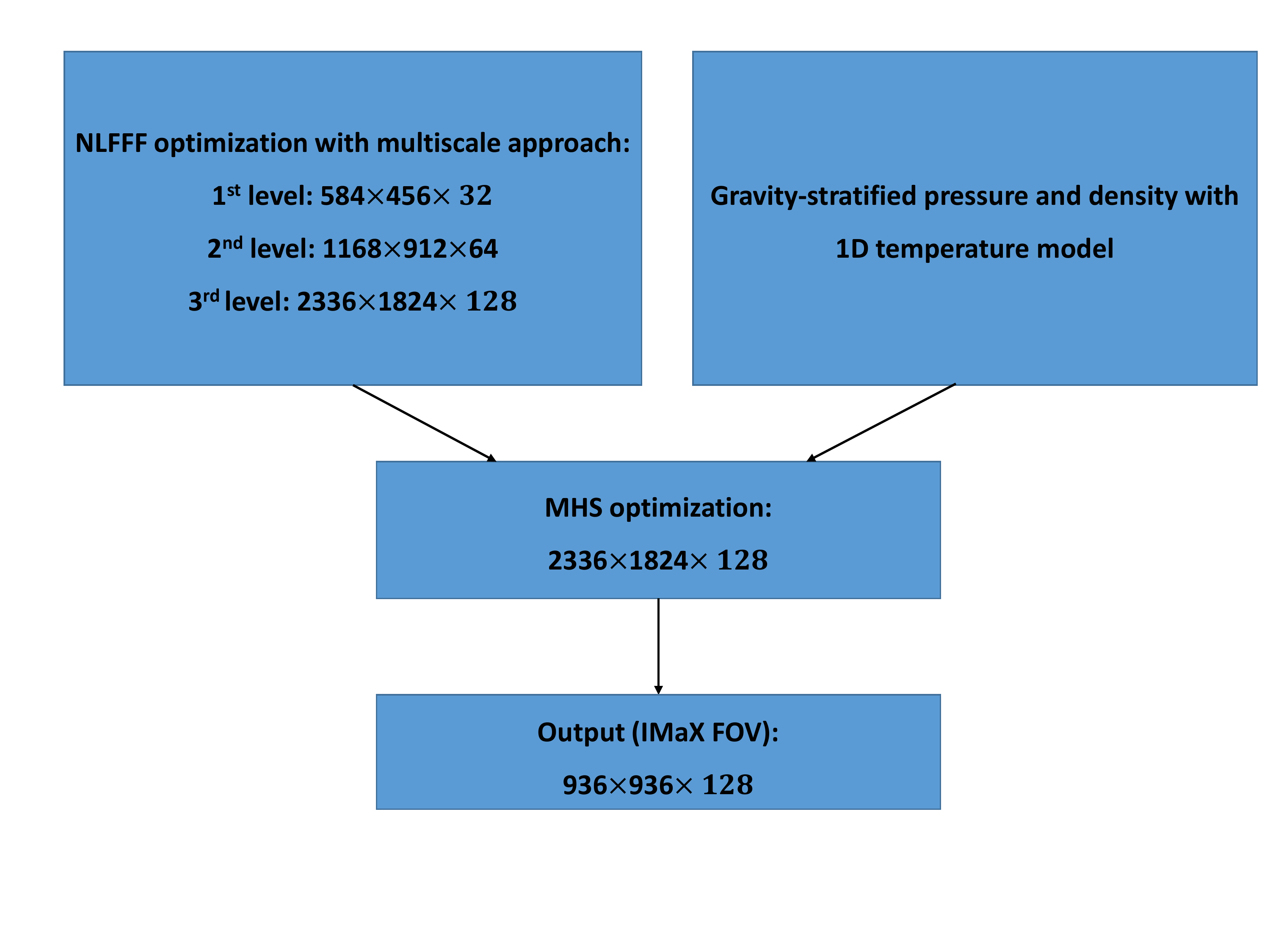}
\caption{Schematic flow chart of the MHS code applied to a vector magnetogram.} 
\label{fig:flowchart}
\end{figure}

\begin{figure}
\includegraphics[width=\hsize]{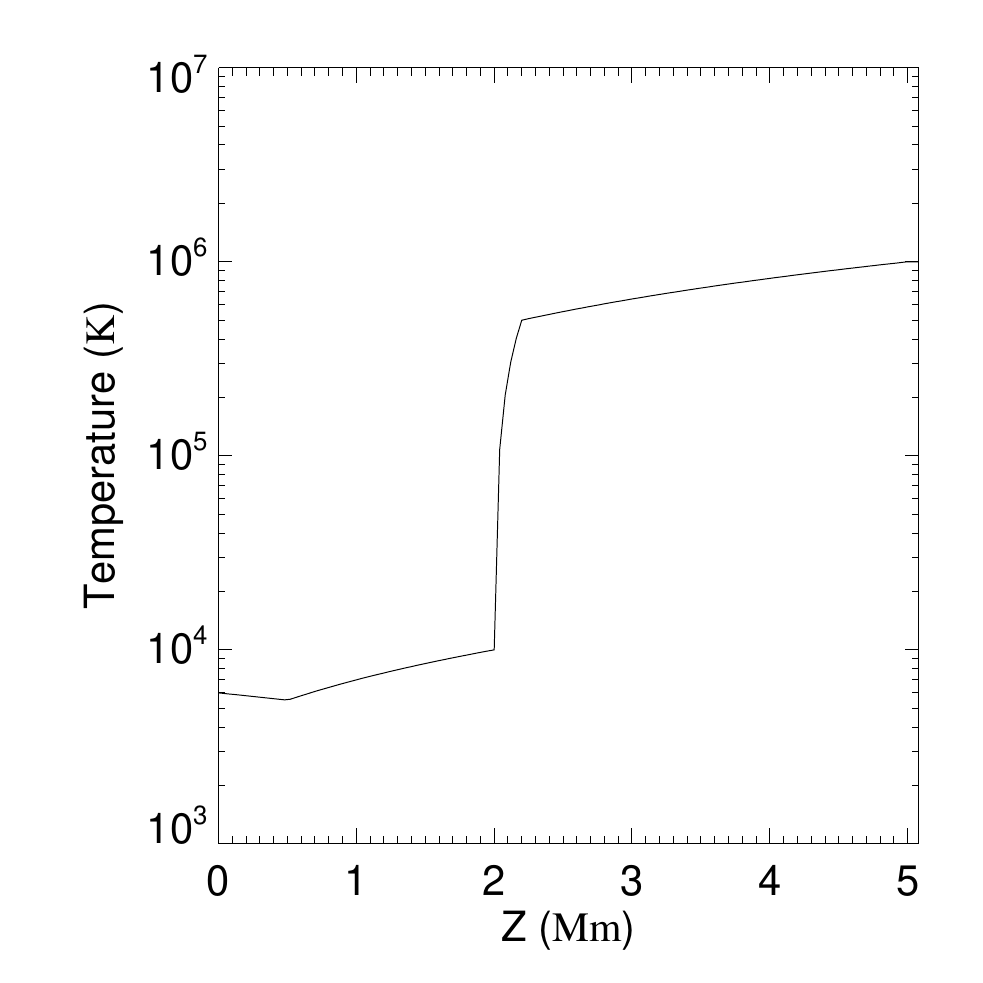}
\caption{Temperature profile of the gravity-stratified atmosphere employed as the initial condition.}
\label{fig:temperature_1D}
\end{figure}

Fig.~\ref{fig:flowchart} depicts the flow chart of our code applied to the combined magnetogram. The following steps are plotted. First, compute a NLFFF \citep{w04} with a preprocessed magnetogram \citep[the net Lorentz force and torque are removed within the error margin of the measurement by using a minimization principle to make the data compatible with the force-free assumption]{wis06} and a gravity-stratified atmosphere along field lines with a 1D temperature profile (Fig.~\ref{fig:temperature_1D}). The pressure on the bottom boundary is computed using $p=p_{quiet}-\frac{1}{2}{B_z}^2$, where $p_{quiet}$ is the pressure in the quiet region. The bottom density is determined by assuming a uniform temperature of $6000$ Kelvin on the photosphere. Second, carry out a further optimization to achieve an MHS equilibrium with the original magnetogram. Note that although a temperature profile is given at the initial state to relate the gas pressure and density, the initial temperature profile is no longer a restriction on the pressure and density optimization. There are two options to input the bottom magnetic boundary. One is to replace the preprocessed magnetogram with the original magnetogram directly, which is used in this study. The other is to change the magnetic field gradually from the preprocessed value to the original value. The side and top boundaries are fixed to their initial values which are potential fields. The computation is performed in a $2336\times1824\times128$ box with a 40 km grid spacing both in horizontal and vertical directions. All of the following analyses are restricted to a $936\times936\times128$ box above the IMaX FOV (unless otherwise stated).
      
\section{Analysis of the extrapolation results} \label{sec:results}

\subsection{Solution consistency}

\begin{figure}
\includegraphics[width=\hsize]{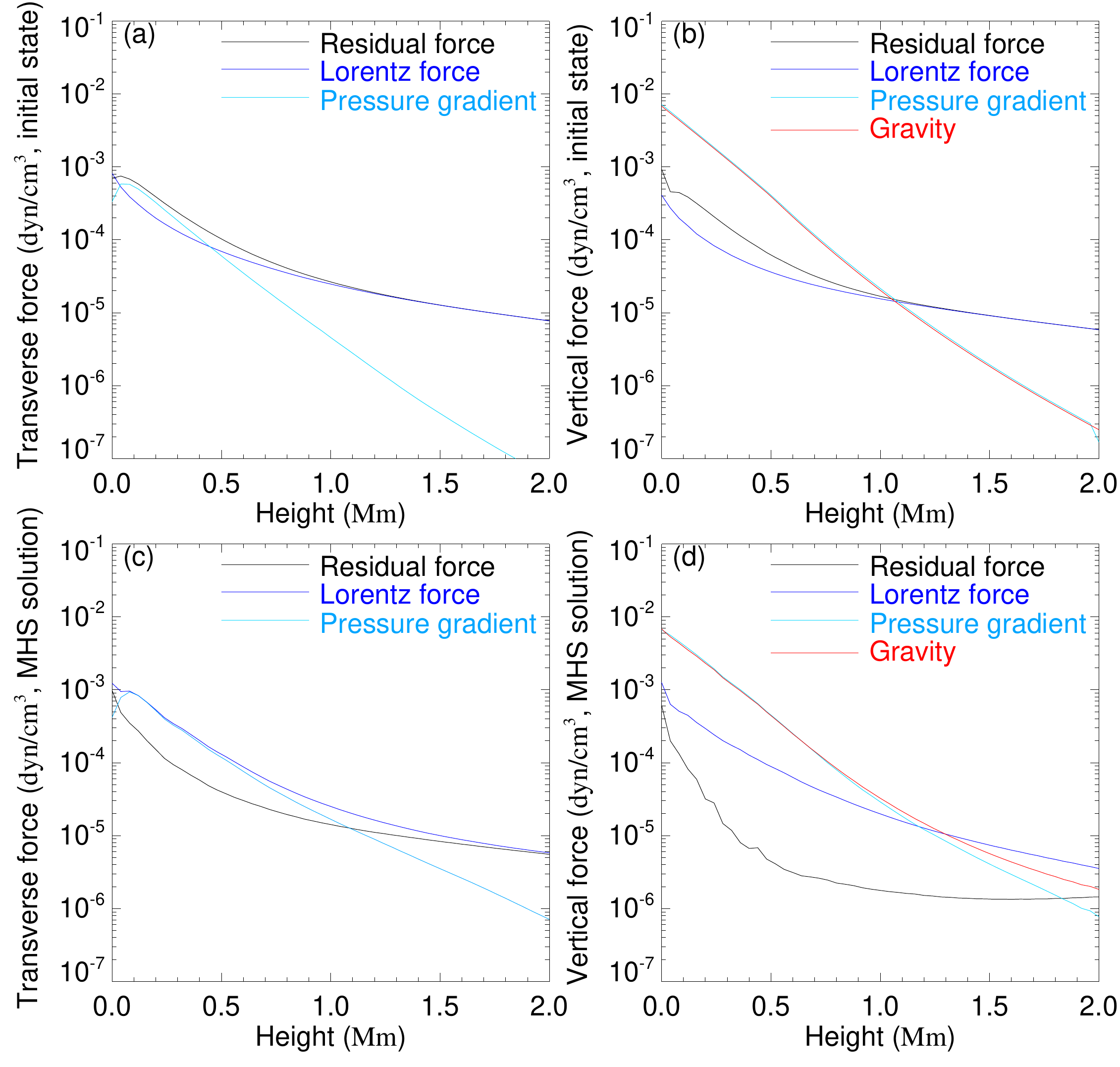}
\caption{Planar average of forces as a function of height in the initial unrelaxed state (a)(b) and in the MHS equilibrium (c)(d). Panels (a) and (c) show the transverse direction, while panels (b) and (d) illustrate the forces in the vertical direction.}
\label{fig:residual_force}
\end{figure}

Both the magnetogram and the disequilibrium of the initial atmosphere (due to the nonuniform gas pressure and density in the photosphere) drive the evolution of the system to an MHS state. Fig.~\ref{fig:residual_force} illustrates the compensation of forces in the initial state and in the final equilibrium. The horizontal component of the forces are shown in panels (a) and (c), while the vertical components are illustrated in panels (b) and (d). In the initial state, we find the residual force is greater than the Lorentz force (see panels (a) and (b)), so that this state is obviously far from an MHS equilibrium. As mentioned before, this promotes further optimization. We see, in the final solution, that the Lorentz force is balanced effectively by the pressure gradient and vertically also by gravity (see panels (c) and (d)). On average, the residual force is 43\% of the Lorentz force in the transverse direction while the percentage is only 24\% in the vertical direction below 2 Mm. These residual forces are not close to 0. The main contributions come from the photosphere and regions with very high or low $\beta$. The inadequate boundary condition of plasma as well as the noise in the measured magnetic field prevent the MHS extrapolation from balancing forces well on the photosphere. In a very high plasma region ($\beta > 10$) Lorentz force are too weak to act against plasma forces, which could result in a ratio of residual force over Lorentz force that is much larger than 1. In a very low plasma region ($\beta < 0.1$) plasma forces are too weak to against Lorentz force, which could result in a ratio that is close to 1. We recompute the ratios excluding the bottom boundary and only in regions where $0.1 \geq \beta \leq 10$. The numbers are 23\% and 7.6\% for the transverse and vertical direction, respectively, which means that the major part of the Lorentz force is balanced. Although those two numbers are still not very small, they are acceptable considering that we have embedded two data sets in each other (recorded by different instruments and obtained by different inversion technologies, etc.) when producing the magnetogram that provides the bottom boundary.

\subsection{Plasma}

\begin{figure}
\includegraphics[width=\hsize]{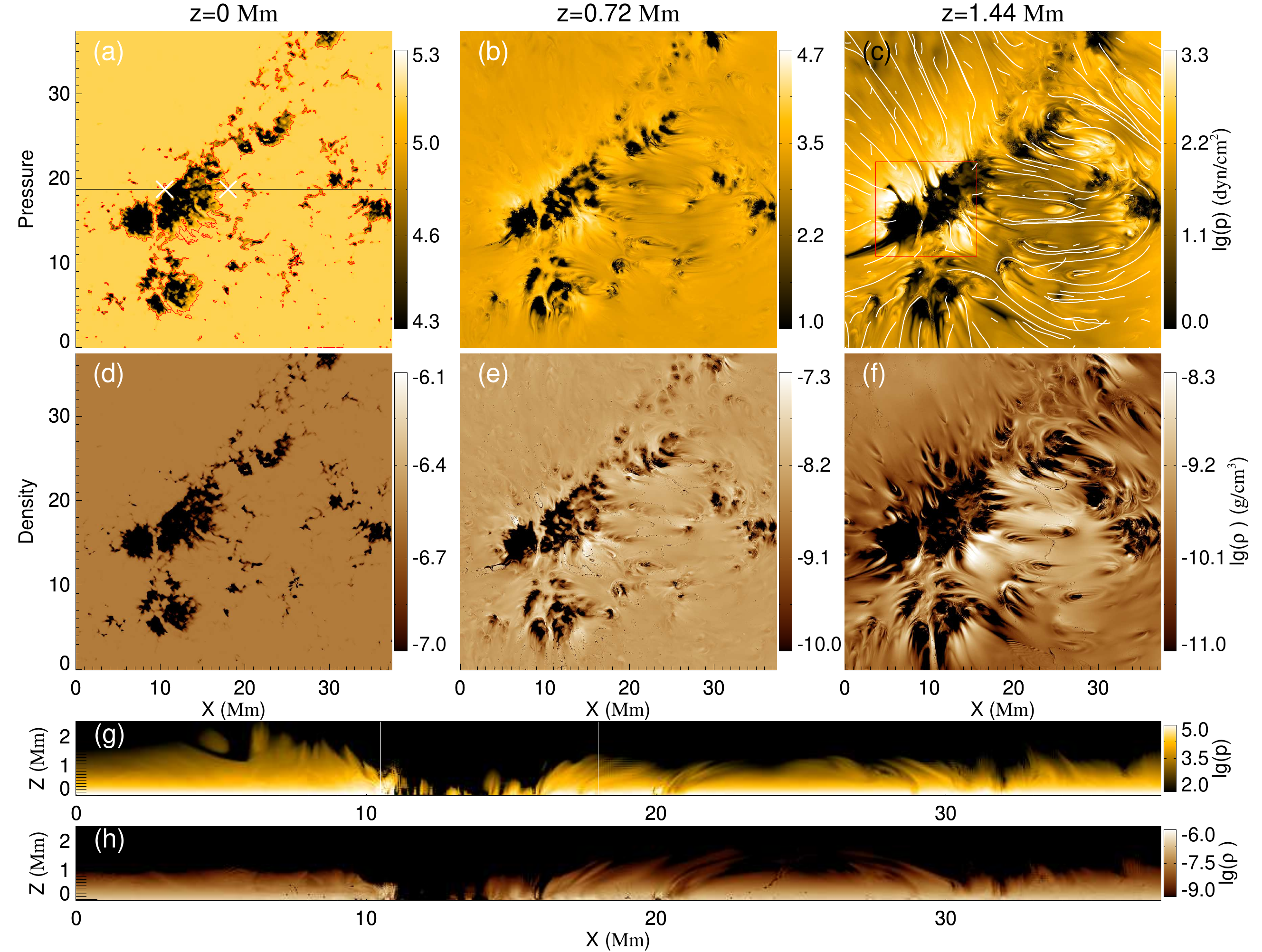}
\caption{(a)-(f) Gas pressure and density at heights of 0 Mm, 0.72 Mm and 1.44 Mm. (g)(h) Pressure and density at the plane y=18.7 Mm (black line in panel (a)). The red contours in panel (a) outline the locations at which magnetic field strength is 1000 G. Two white ``X'' in panel (a) indicate the intersections of the 1000 G contour and the black line. They are also the seed points of the two white lines in panel (g). The red rectangle in panel (c) specifies the FOV of Fig.~\ref{fig:squeeze}. Uniformly selected magnetic field lines in panel (c) range from 600 km to 1400 km.}
\label{fig:pressure_density_cut}
\end{figure}

\begin{figure}
\includegraphics[width=\hsize]{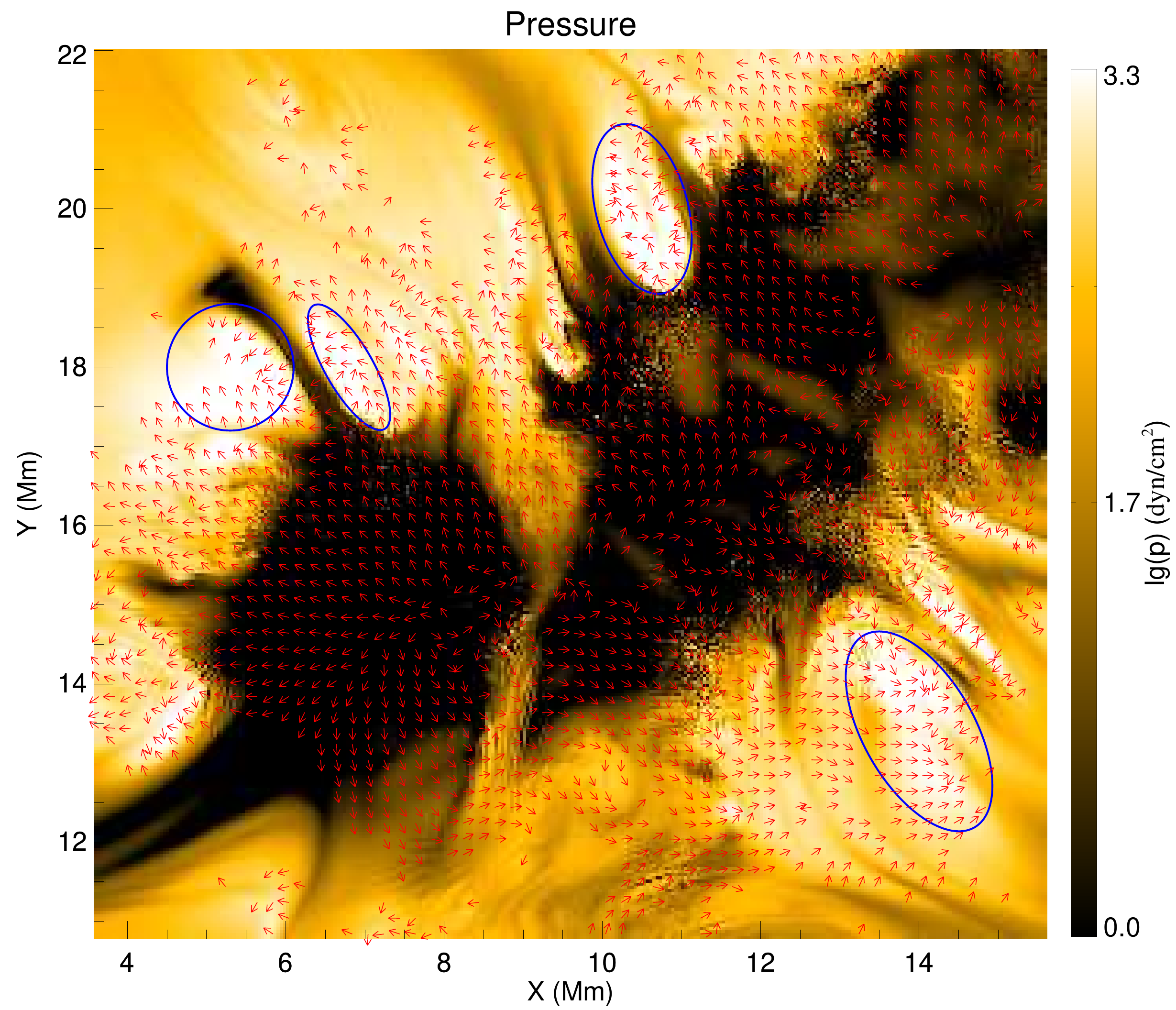}
\caption{Two-dimensional Lorentz force vectors overplotted on the gas pressure at a height of 1.44 Mm. The plotted FOV is outlined by the red rectangle in Fig.\ref{fig:pressure_density_cut} (c). Blue ellipses encircle four typical regions where the gas pressure is enhanced.}
\label{fig:squeeze}
\end{figure}

\begin{figure}
\centering
\includegraphics[width=\hsize]{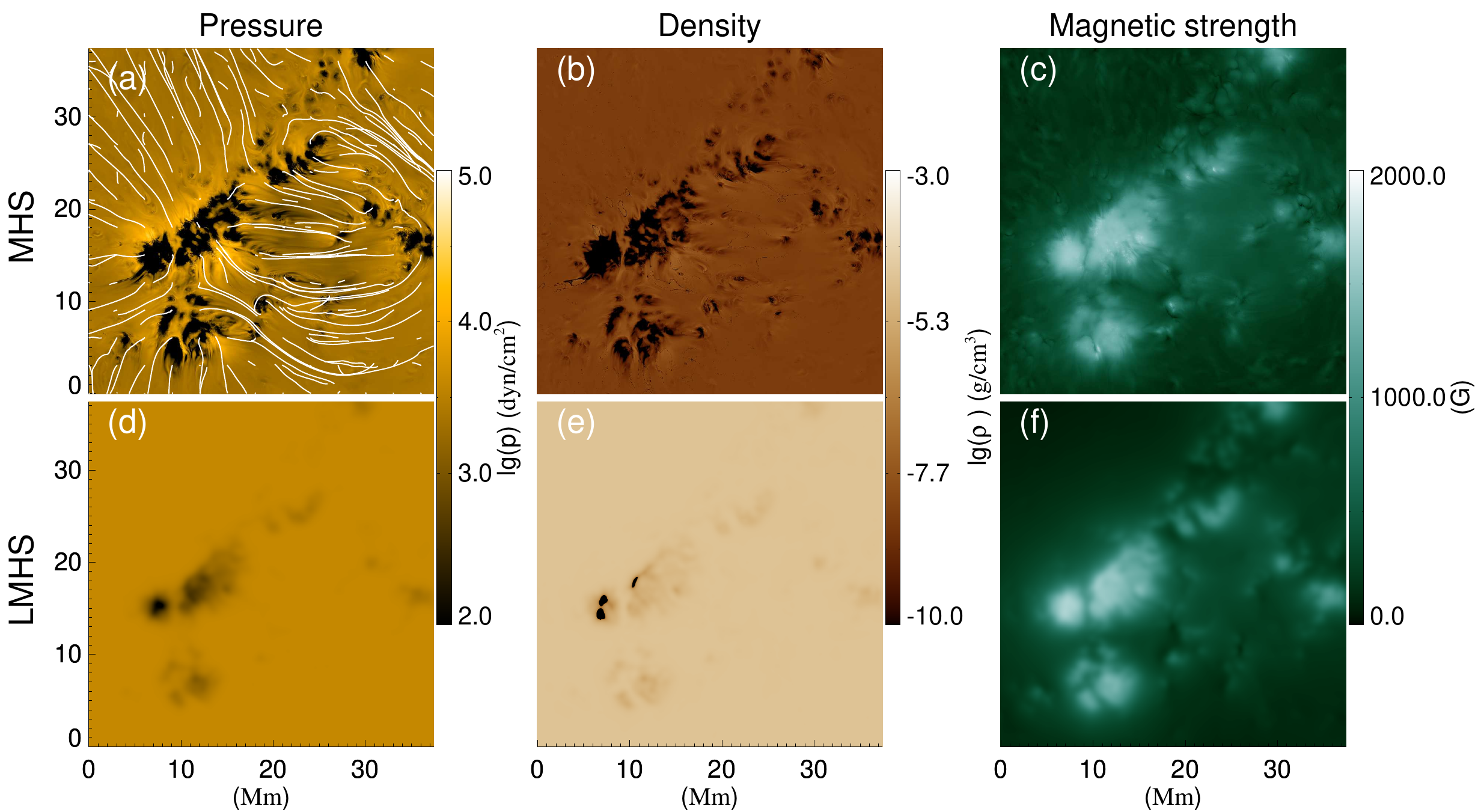}
\caption{Pressure, density and magnetic field strength of the MHS model (a)(b)(c) and the LMHS model (d)(e)(f) at a height of 0.72 Mm. Uniformly selected magnetic field lines in panel (a) range from 600 km to 1400 km.}
\label{fig:plasma_compare}
\end{figure}

Fig.~\ref{fig:pressure_density_cut} shows plasma distributions from different perspectives. From both top and side views, we see clearly that plasma pressure and density are reduced in strong field regions to keep the force balance. This is consistent with sunspot observations. Similar results were also found both in the LMHS modelings \citep{ads98,adm99,wnn15,wnn17} and the previous tests of our MHS code \citep{zw18,zw19}. However, we also find that, at some parts of the active region edges, plasma pressure and density are enhanced. This has never been reported in previous extrapolations. Fig.~\ref{fig:squeeze} depicts selected Lorentz force vectors in a FOV that is outlined by the red square in Fig.~\ref{fig:pressure_density_cut} (c). We see that regions with an enhanced gas pressure which are encircled by blue ellipses are surrounded with a net inward flux of the Lorentz force. As a result, the plasma is squeezed together. It is also worth noting that the fibril-like plasma pattern traces the magnetic field in Fig.~\ref{fig:pressure_density_cut} (c) and Fig.~\ref{fig:plasma_compare} (a). Such localized concentrations reflect the nonlinear nature of the MHS system, which cannot be observed in the LMHS model (see Fig.~\ref{fig:plasma_compare}).

\begin{figure}
\includegraphics[width=\hsize]{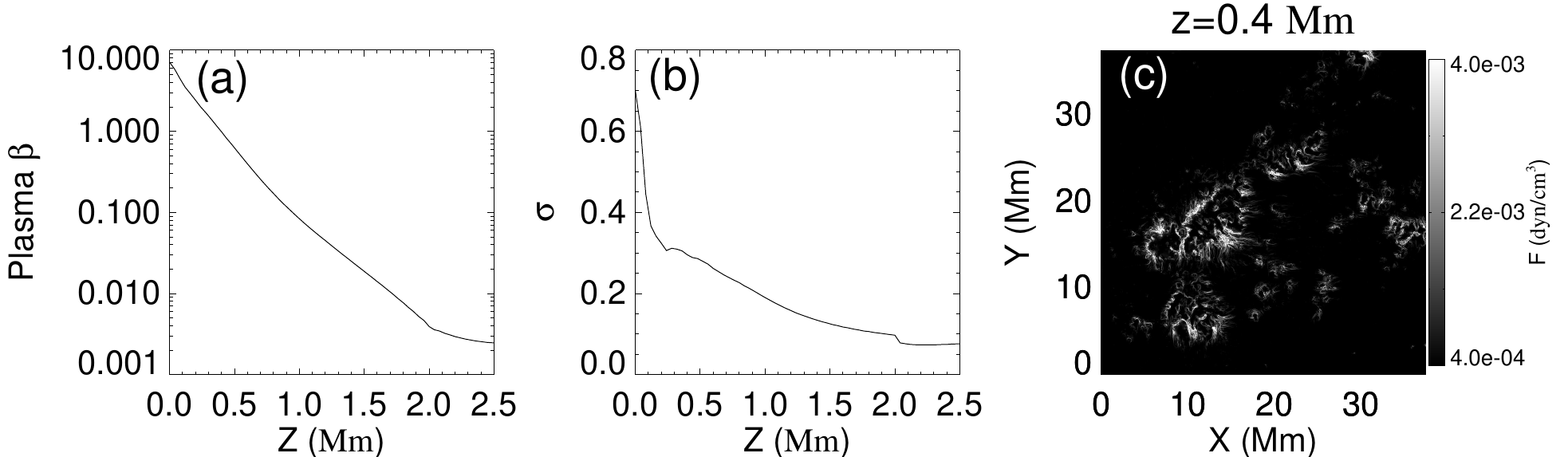}
\caption{Planar average of plasma $\beta$ (a) and of current-weighted sine $\sigma$ (b), where $\sigma$ is the angle between the magnetic field and current, as functions of height. (c) The magnitude of Lorentz force at a height of 0.4 Mm.}
\label{fig:LorentzForce}
\end{figure}

According to the model of plasma $\beta$ over an active region developed by \cite{g01}, the magnetic field dominates plasma above a height of $1\sim 2$ Mm (see Fig.~3 in \cite{g01}). Fig.~\ref{fig:LorentzForce} (a) shows that the $\beta_{(z)}$ of the MHS equilibrium is right inside the $\beta$ range illustrated in \cite{g01}. A high plasma $\beta$ is a necessary, but not a sufficient condition for the magnetic field to be non-force-free. To see whether the magnetic field is non-force-free in the high $\beta$ region, the current-weighted sine of the angle between the magnetic field and the electrical current density \citep{sdm06} is computed
\begin{equation}
\sigma=\left.\displaystyle\sum_i{\frac{|\mathbf J_i\times \mathbf B_i|}{B_i}} \right/ \sum{J_i}.
\label{eq:cwsin}
\end{equation} As shown in Fig.~\ref{fig:LorentzForce} (b), $\sigma$ decreases fast from 0.7 at the photosphere to less than 0.1 above 2.0 Mm. Since the effective vertical resolution of the solution is roughly the same as the horizontal resolution of the magnetogram, the high resolution IMaX data allow us to study this non-force-free layer in detail. Coarser data (e.g. HMI) with a few grid points to resolve this layer meet with difficulties when focussing on the lower atmopshere. Fig.~\ref{fig:LorentzForce} (c) shows Lorentz force distributions at a height of 0.4 Mm. We see strong Lorentz forces are mainly located at edges of strong-field features, where plasma $\beta$ drops precipitous. This is a natural result caused by the strong plasma forces at edges. The great plasma differences at edges of magnetic features in the lower atmosphere are routinely observed.

\subsection{Relation between plasma pressure, current density and photospheric brightness}

\begin{figure}
\centering
\includegraphics[width=\hsize]{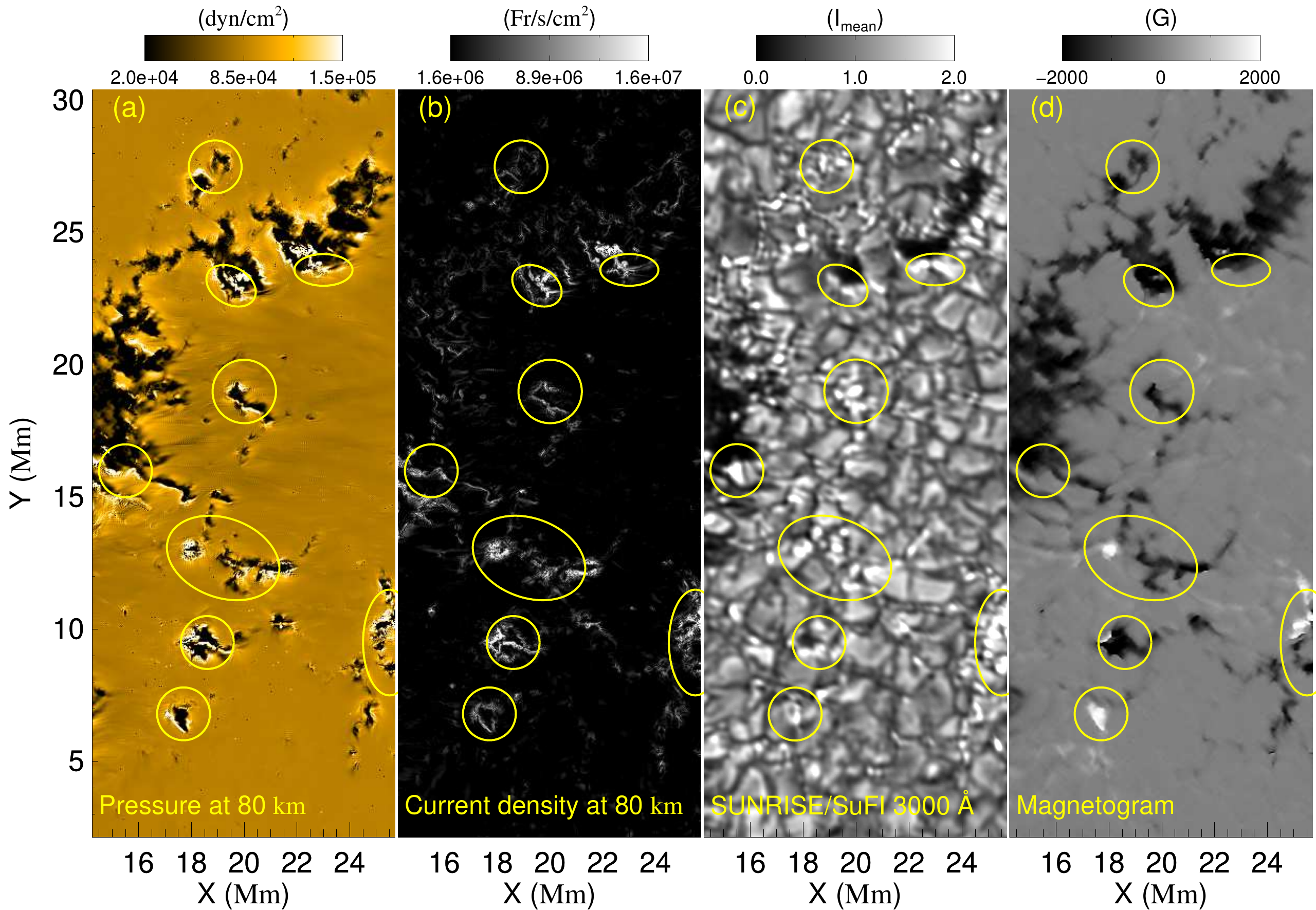}
\caption{(a) Gas pressure and (b) current density at height of 80 km. (c) SuFI 3000 $\AA$ image. (d) IMaX magnetogram. Strong pressure regions are outlined by yellow ellipses.}
\label{fig:pre_sufi300}
\end{figure}

Fig.~\ref{fig:pre_sufi300} shows the mapping of plasma and current density onto an image acquired by \textsc{Sunrise}/SuFI. Note that the extrapolation data are cut according to the SuFI FOV of $15''\times 38''$. Bright points are clearly seen in the inter-granular lanes in panel (c). They are typically regarded as nearly vertical slender flux tubes with kG magnetic fields \citep{s93,nts08,rdb14}. The lateral radiative inflow makes the tube hot and bright \citep{s76}, making them visible as bright points at wavelengths sensitive to temperature \citep{ssb03,rsm10}.

Fig.~\ref{fig:pre_sufi300} (a) (b) show that regions of high plasma pressure and strong electric current density coincide. Most of them are located near the edges of magnetic flux tubes. These flux tubes which appear as photospheric bright points at SuFI 3000 $\AA$ (see Fig.~\ref{fig:pre_sufi300} (c)) are often accompanied by high plasma pressure (below 400 km) and electric current around them. At the edges of flux tubes, the plasma and magnetic field interact, which leads to the high current and co-spatial high plasma pressure. Such a high current density around magnetic bright points is also reminiscent of the electric current sheets expected to bound flux tubes. It is worth noting that many bright points do not have corresponding enhancements in the electric current. This may be related to the local dynamics. It must also be kept in mind that the MHS model does not take into account the radiation.

\subsection{Comparisons of magnetic fields and chromospheric fibrils}

\begin{figure}
\centering
\includegraphics[width=\hsize]{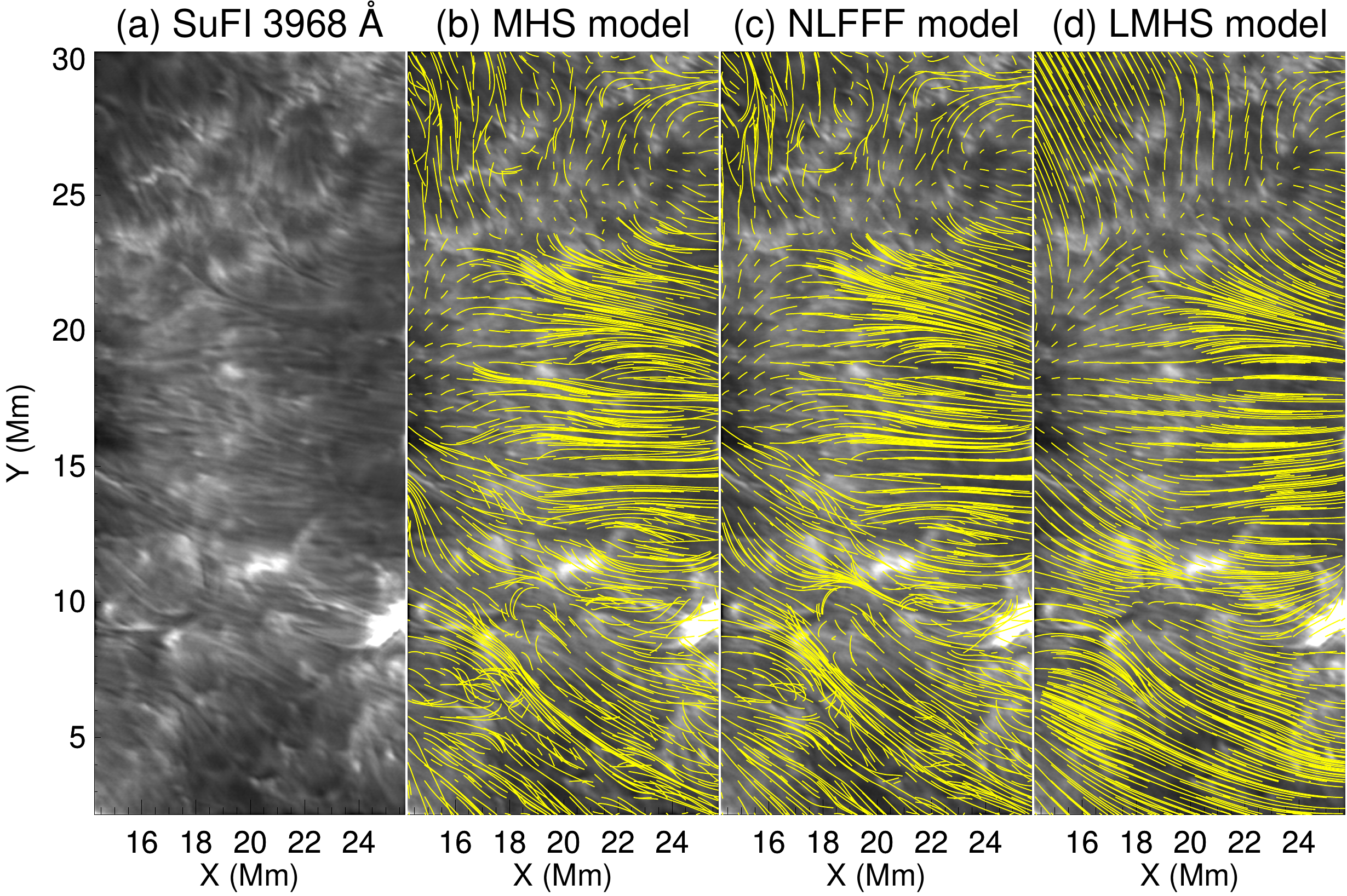}
\caption{Magnetic field lines of different models within the heights [600, 1400] km overplotted on the image observed in Ca II H core line with 1.1 $\AA$ wide filter. The field lines are rooted in the same set of seed points in all panels.}
\label{fig:fl_sufi397}
\end{figure}

The SuFI instrument provides diffraction-limited images at 3968 $\AA$ with contributions from both the photosphere and low-to-mid-chromosphere \citep{jss17}. The observed slender fibrils at this wavelength are the dominant structures (see Fig.~\ref{fig:fl_sufi397} (a)) in the SuFI FOV. It is generally believed that long fibrils in the chromosphere outline magnetic fields in this layer \citep{ds11,jyr11,spl13,lcr15,zwd16}. We plot field lines within the sub-volume spanning the 600-1400 km height range, which implies that low field lines ($\leq$ 600 km) are ignored. Seed points of the field lines are uniformly selected in the photosphere. Fig.~\ref{fig:fl_sufi397} shows that most fibrils trace field lines nicely. The similar field line patterns of different extrapolation models imply that plasma forces have limited impact on the large scale magnetic field of this potential-like active region, at least in the height range from 600-1400 km.

In order to quantitatively show the degree of agreement between fibrils and field lines we compute the angle $\theta$ between fibrils and the plane-of-the-sky component of the magnetic field. The computation is not carried out on every pixel on the image. Instead, to show more clearly the discrepancy among different models, we focus on the regions where large differences between magnetic vectors of the MHS model and the NLFFF model appear (see contours in Fig.~\ref{fig:fl_sufi397}). We have tried to identify all fibrils in the regions of interest. For each fibril, one point is picked for the statistics. Then the method introduced by \citet{ifw08} is used to estimate the orientation of the fibril. The key point of the method is to use the gradient and Hessian matrix of the image intensity to determine the orientation. Then $\theta$ is computed vertically within the 600-1400 km height range. The smallest one is specified as the discrepancy between the fibril and the magnetic vector. For totally 26 points picked (hence 26 fibrils identified), the average of $\theta$ for the MHS model, NLFFF model and LMHS model are $11.8^\circ,\ 15.7^\circ$ and $20.9^\circ$, respectively. There are 18 points (nearly 70\%) at which the MHS model's $\theta$ is less than those of the other two models.

\begin{figure}
\centering
\includegraphics[width=\hsize/2]{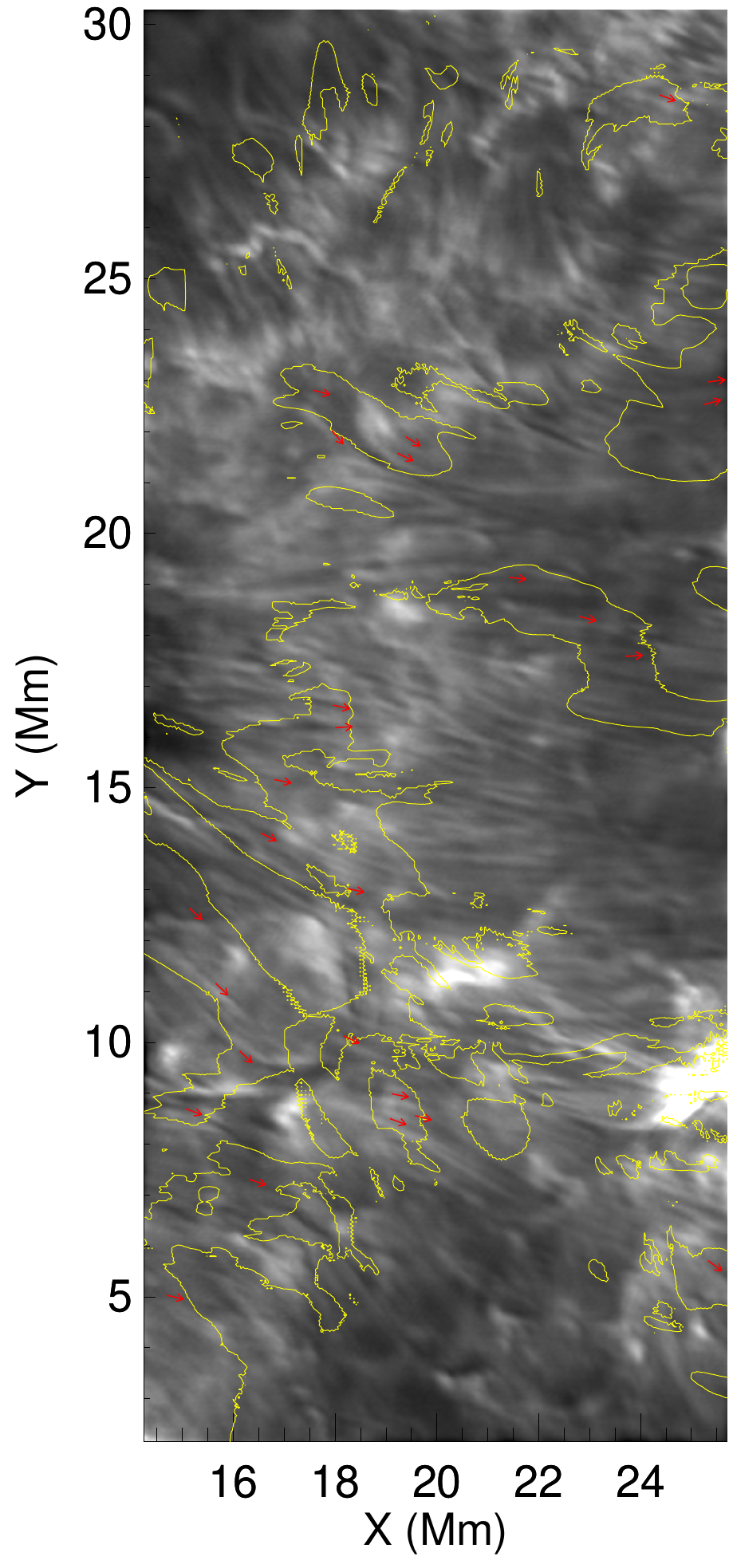}
\caption{Contours show where the average of angles between the MHS and NLFFF's magnetic vectors range from 600-1400 km equals to $5^\circ$. Red vectors depict orientations of the fibrils. Each vector is started from the point at which $\theta$ has the largest value over this fibril.}
\label{fig:vec_sufi397}
\end{figure}

\section{Conclusion and perspectives} \label{sec:conclusion}

In this work we apply, for the first time, a nonlinear MHS code to model the solar lower atmosphere starting from a real magnetogram. A combined vector magnetogram to cover the whole active region is used as the boundary input.

MHS equilibrium is constructed in which Lorentz forces are effectively balanced by plasma forces. The pressure and density depletion take place in the strong field regions together with enhancements in the active region edges. In the low $\beta$ layers, the fibril-like plasma structures clearly outline the magnetic field lines. A thin non-force-free layer is resolved within about 50 grid points in the z direction. The high plasma pressure and co-spatial high electric current appear around the bright points prominent in SuFI 3000 $\AA$ images. Although similar general patterns of the magnetic fields are found in different types of extrapolations (MHS, linear MHS, nonlinear force-free fields), a quantitative comparison implies that the magnetic field vectors of the MHS model are more aligned with the orientation of the chromospheric fibrils observed at SuFI 3968 $\AA$.

As mentioned before, the gas pressure and density are optimized independently. That means the initial isothermal temperature profile can not be kept. An equation of state is needed to close the MHS equations. However, the present version of the code deals with pressure and density separately. Future studies may focus on the extrapolation under more constrains from observations.

The MHS extrapolation converges at a much lower speed than the NLFFF extrapolation. In this application with $2336\times1824\times128$ grid points, the MHS code runs about 55 hours on 2 Intel Xeon Gold 6150 CPU with 18 cores while the NLFFF code only takes 1 hour. Considering the thinness of the non-force-free layer, a future extrapolation could consist of a time-consuming MHS model in the lower atmosphere and a much faster NLFFF model in the higher atmosphere by using the model below as the boundary input. Such a combination is a superior solution to the NLFFF model with preprocessed magnetogram \citep{wis06}. Last but not least, the MHS equilibrium, as computed by the new model, can serve as the initial conditions for time-dependent data-driven MHD simulations \citep{jwf16,gxk19}.





\begin{acknowledgements}
We acknowledge constructive suggestions of the referee and valuable discussions with B. Inhester and T. Riethm\"{u}ller. The German contribution to \textsc{Sunrise} and its reflight was funded by the Max Planck Foundation, the Strategic Innovations Fund of the President of the Max Planck Society (MPG), DLR, and private donations by supporting members of the Max Planck Society, which is gratefully acknowledged. This project has also received funding from the European Research Council (ERC) under the European Union’s Horizon 2020 research and innovation programme (grant agreement No. 695075) and has been supported by the BK21 plus program through the National Research Foundation (NRF) funded by the Ministry of Education of Korea. This work was also supported by DFG-grant WI 3211/4-1.
\end{acknowledgements}

%
   \bibliographystyle{aa} 
   \bibliography{2019} 
%

\end{document}